\documentclass{article}

\usepackage{PRIMEarxiv}

\usepackage[utf8]{inputenc} 
\usepackage[T1]{fontenc}    
\usepackage{hyperref}       
\usepackage{url}            
\usepackage{booktabs}       
\usepackage{amsfonts}       
\usepackage{nicefrac}       
\usepackage{microtype}      
\usepackage{lipsum}
\usepackage{amsmath}
\usepackage{amssymb}
\usepackage{fancyhdr}       
\usepackage{graphicx}       
\graphicspath{{media/}}     
\usepackage{braket}

\usepackage{siunitx}  

\usepackage{caption}
\title{Unlocking the Potential of Photoexcited Molecular Electron Spins for Room Temperature Quantum Information Processing}

\author{
  Kuan-Cheng Chen  \\
  Department of Materials \\
  Imperial College London \\
  South Kensington, London, SW7 2AZ, United Kingdom\\
   \And
  Alberto Collauto \\
  Department of Chemistry and Centre for Pulse EPR spectroscopy (PEPR)\\
  Imperial College London \\
  Molecular Sciences Research Hub\\
  London, W12 0BZ, United Kingdom\\
    \And
  Ciarán J. Rogers \\
  Department of Chemistry and Centre for Pulse EPR spectroscopy (PEPR)\\
  Imperial College London \\
  Molecular Sciences Research Hub\\
  London, W12 0BZ, United Kingdom\\
  \AND
  Shang Yu \\
  Blackett Laboratory, Department of Physics,\\
  Imperial College London \\
  South Kensington, London, SW7 2BW, United Kingdom\\
   \AND
  Mark Oxborrow  \\
  Department of Materials \\
  Imperial College London \\
  South Kensington, London, SW7 2AZ, United Kingdom\\
   \And
  Max Attwood\thanks{\textit{Corresponding Author}: m.attwood@imperial.ac.uk
  }  \\
  Department of Materials \\
  Imperial College London\\
  South Kensington, London, SW7 2AZ, United Kingdom\\
  \texttt{m.attwood@imperial.ac.uk} \\
}

\begin{document}
\maketitle

\begin{abstract}
Future information processing technologies like quantum memory devices have the potential to store and transfer quantum states to enable quantum computing and networking. A central consideration in practical applications for such devices is the nature of the light-matter interface which determines the storage state density and efficiency. Here, we employ an organic radical, $\alpha$,$\gamma$-bisdiphenylene-$\beta$-phenylallyl (BDPA) doped into an o-terphenyl host to explore the potential for using tuneable and high-performance molecular media in microwave-based quantum applications. We demonstrate that this radical system exhibits millisecond-long spin-lattice relaxation and microsecond-long phase memory times at room temperature, while also having the capability to generate an oscillating spin-polarized state using a co-dissolved photo-activated tetraphenylporphyrin moiety, all enabled by using a viscous liquid host. This latest system builds upon collective wisdom from previous molecules-for-quantum literature by combining careful host matrix selection, with dynamical decoupling, and photoexcited triplet-radical spin polarisation to realise a versatile and robust quantum spin medium.
\\

\end{abstract}


\section{Introduction}

The essence of quantum technology lies in the ability to exert coherent control over two or more energy levels, corresponding to qubits and  qudits, respectively. Across coherent quantum computing platforms, such as superconducting circuits\cite{clarke2008superconducting} and silicon spin qubits\cite{maurand2016cmos}, cryogenic conditions are required to mitigate the effects of environmental thermal noise (associated with qubit fidelity\cite{huang2024high}) and to enhance the efficacy of quantum operations. There are two approaches to improve the efficacy of quantum operations: one involves reducing the quantum gate error rate through the pulsed protocol of quantum gates\cite{souza2012robust}, and the other involves enhancing the material’s inherent coherence and decoherence times through material design\cite{zadrozny2015millisecond,atzori2018structural,bayliss2022enhancing,amdur2022chemical}.

\begin{figure}[htpb]
  \centering
  \includegraphics[width=0.7\linewidth]{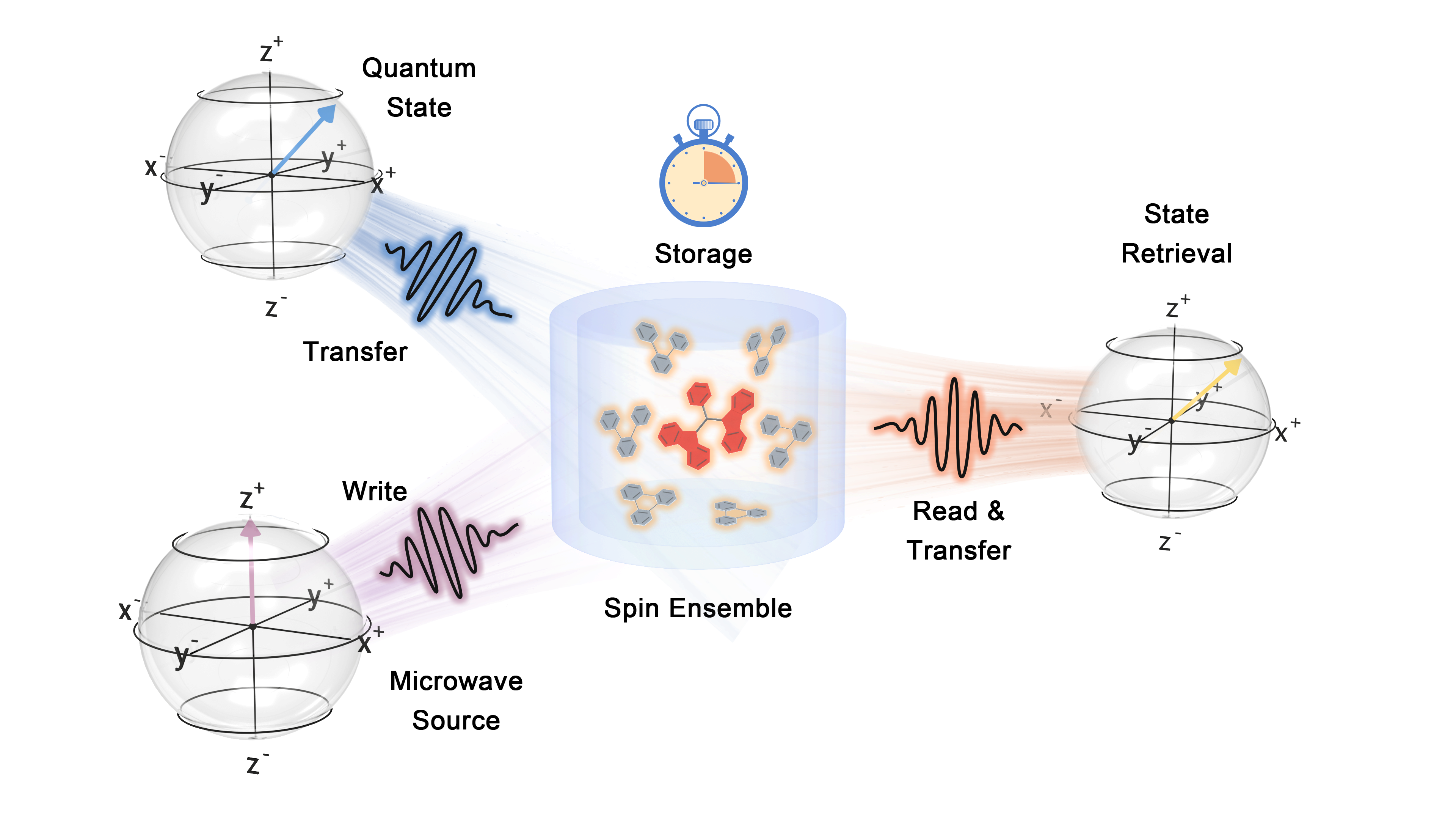}
  \caption{Conceptual diagram of molecular spin ensembles for microwave storage and quantum memory.}
  \label{fig:fig1}
\end{figure}

From the perspective of material design\cite{basov2017towards}, particularly for materials relying on electronic spins, the primary objective is to obtain long spin-lattice relaxation time (\(T_1\)) and long phase memory times (\(T_m\)) such that spins can be manipulated and measured without undesirable spin decay associated with processing errors and loss of sensitivity\cite{Polash2023}. Further practical caveats become important for particular applications. For example, in quantum sensing and information processing it is necessary to initiate a spin-polarised quantum state usually via optical excitation before spin manipulation and readout\cite{sushkov2014all, degen2017quantum}. While spin manipulation is most commonly performed using microwave pulses, the readout mechanism can be based on microwave or optical regimes with the latter being generally desirable because it is instantaneous. For quantum memory (Fig.~\ref{fig:fig1}), the storage efficiency of a quantum state or microwave pulse is determined by the coupling strength between a superconducting circuit or microwave resonator and a spin ensemble, which scales by $\sqrt{N}$, where $\textit{N}$ is the number of spins\cite{Kubo2012}. However, the storage time is determined by the spin dephasing time (\(T_2^*)\) and in randomly doped solid-state materials \(T_2^*\) is proportional to 1/$\textit{N}$. These contrasting performance relationships make it particularly challenging to engineer a family of quantum materials for diverse applications, especially since \(T_2^*\) is challenging to measure precisely without an optical readout \cite{de2021materials}.

In this report, we reveal our initial explorations using stable organic radicals to generate robust RT quantum technologies, and in particular focus here on their potential application in memory. By utilizing the well-known organic radical, $\alpha$,$\gamma$-bisdiphenylene-$\beta$-phenylallyl (BDPA) doped into an organic host, o-terphenyl (OTP), we demonstrate a long and temperature-insensitive \(T_m\) reaching 2.33 \(\mu\)s at room temperature (RT) with a relatively high molar concentration of 0.1$\%$. The effective coherence time was significantly lengthened using a dynamical decoupling approach to 11.5 \(\mu\)s. It is shown that the temperature insensitivity is due to dilution in the OTP host, which becomes a metastable and optically transparent viscous liquid after melting and cooling to RT. We finally demonstrate that the long \(T_1\) associated with BDPA can be used in conjunction with photo-activated triplets to realize record long-lived and even aberrant modes of spin polarisation. 


\section{Experimental Results}
\subsection{Investigation of spin coherence by pulsed EPR spectroscopy}

\begin{figure}[htpb]
  \centering
  \includegraphics[width=0.85\linewidth]{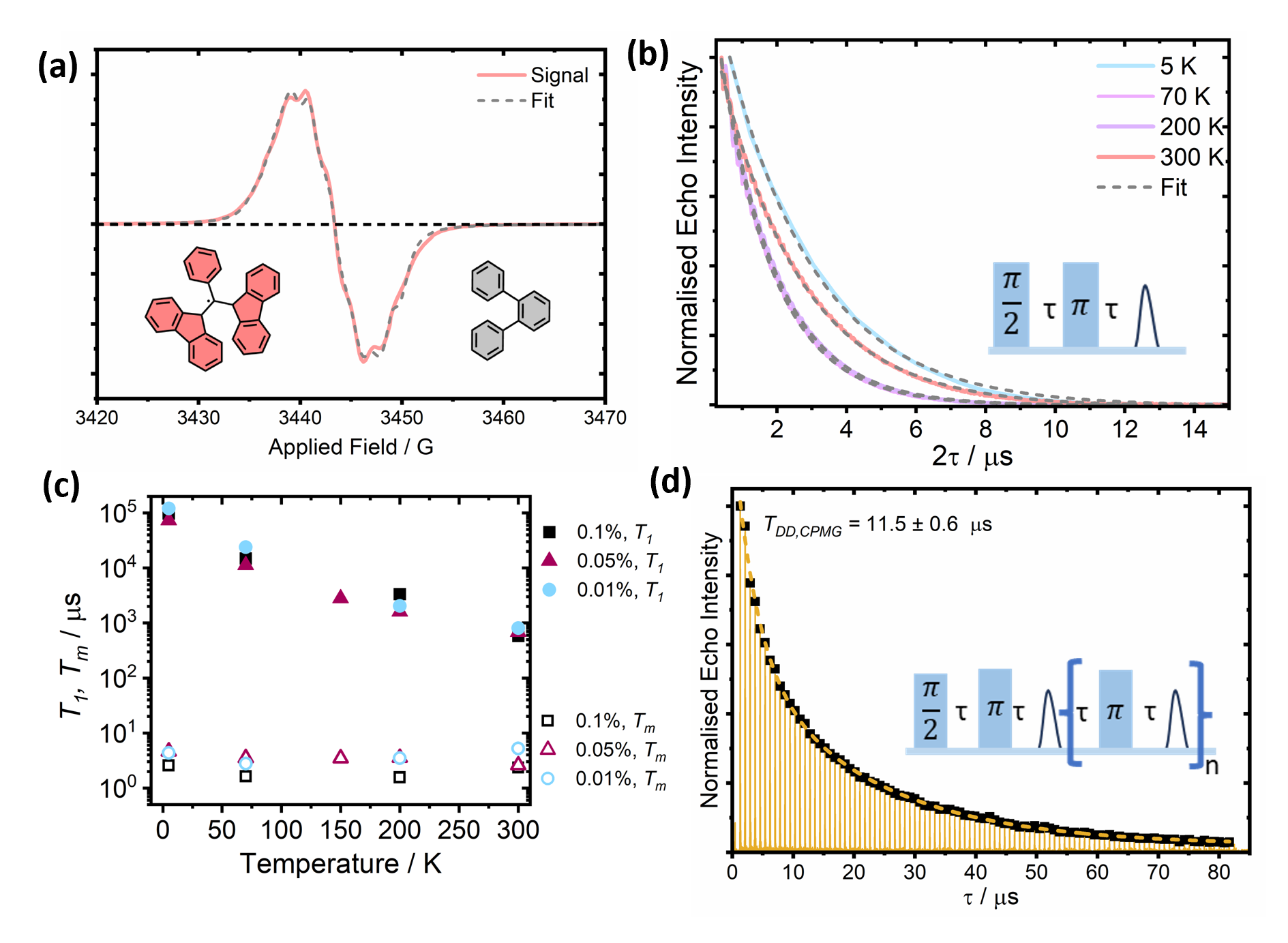}
  \caption{\label{Fig:IRandHahn}(a) Steady-state EPR signal at RT for BDPA in OTP, with S=1/2 fit shown by the grey dashed line;  (b) Pulsed EPR measurements of the Hahn echo decay for 0.1\%BDPA in OTP; (c) Summary plot showing the variation of \(T_1\) and \(T_m\) over different temperatures and spin concentrations; (d) Carr-Purcell-Meiboom-Gill (CPMG) pulsed sequence with an interpulse delay of 410 ns, overlaid a monoexponential fit.}
  \label{fig:fig2}
\end{figure}

The host OTP was chosen due to its structural similarity to p-terphenyl which acts as a light-stable inert host for aromatic hydrocarbons such as tetracene and pentacene and facilitates robust photoinduced quantum behaviours\cite{Kothe2021, Breeze2017}. However, unlike p-terphenyl, its RT quasi-glass phase can be used as a universal host to accommodate a range of structurally dissimilar and even paramagnetic molecules\cite{Dzuba2006, Dzuba2022}. Dissolution in OTP has previously been used to reduce dipole coupling and successfully increase \(T_m\) in radical graphenoids\cite{Lombardi2022}. Experimentally, we estimate that close to 1\% BDPA is the saturation concentration in OTP, however, to avoid uncertainty in our spin measurements, we restricted the highest concentration to 0.1\%. The CW-EPR spectra of the glass phase 0.1\%BDPA in OTP can be seen in Fig.~\ref{fig:fig2}(a). The centre field signal exhibits a small degree of g-value anisotropy and proton hyperfine coupling that can be approximately fitted assuming four in-equivalent protons, typical for BDPA (see ESI)\cite{Delage-Laurin2021}.

Using pulsed X-band EPR spectroscopy, we performed inversion recovery (\(\pi\)--T--\(\pi/2\)--\(\tau\)--\(\pi\)--\(\tau\)--echo) and Hahn-Echo (\(\pi/2\)–\(\tau\)–\(\pi\)–\(\tau\)–echo) experiments to assess the impact of temperature and dopant concentration on \(T_1\) and \(T_m\), respectively. The biexponential nature of the IR signal was fitted by \(f(t) = A_0 + (1 - A_1 e^{-t/T_r} - A_2 e^{-t/T_s}) \), and indicates distinct rapid and slow relaxation pathways\cite{Yu2020}. In a representative case of a sample doped with 0.1\% BDPA, the weight-averaged \(T_1\) increases from \SI{1.2}{ms} at \SI{300}{K} to \SI{120}{ms} at \SI{5}{K} (Fig.~\ref{fig:fig2}(c)). In contrast, \(T_m\) exhibited a monoexponential decay pattern, indicative of a single dominant decoherence mechanism and demonstrates striking temperature independence (Fig.~\ref{fig:fig2}(b) and (c)). Specifically, \(T_m\) is measured at 2.57 \(\mu\)s at 5 K and 2.33 \(\mu\)s at 300 K for a 0.1\% BDPA concentration. These spin coherence times are remarkably long for an organic radical, and to our knowledge, only previously achieved for BDPA in any medium at 4 K\cite{Bonizzoni2023}, potentially suggesting a protective function of the OTP glass phase which was found to vitrify below ~15\(^{o}C\). Interestingly, while \(T_1\) increases with a reduction in temperature, it exhibits no clear correlation with the concentration of BDPA (see also ESI) suggesting that even at 0.1\% the dipole interactions are minimised. The temperature dependence of \(T_1\) aligns with the Raman spin relaxation process, as described by \(\frac{1}{T_1} = a T^n\) \cite{yu2019concentrated, oanta2022electronic, laorenza2021tunable, chen2016electron}, where \(a\) is a pre-factor, \(T\) represents the temperature under measurement conditions, and \(n\), with a range of 1.5-3, is the exponent characterizing the relationship. This range for \(n\) is notably lower than the commonly observed value of 9, suggesting these measurements are taken above the Debye relaxation temperature. A similar phenomenon has been observed in systems such as 1,3,6,8-tetrakis(4-aminophenyl) pyrene and 3,7-di(4-formylphenyl)-functionalized naphthalene diimide (TAPPy-ND)\cite{oanta2022electronic}, as well as in vanadyl porphyrin\cite{urtizberea2020vanadyl} and vanadyl phthalocyanine (VOPc)\cite{atzori2016room}, where \(n\) is approximately 3. Additionally, at higher temperatures, cross-relaxation likely contributes to spin relaxation, as evidenced by shorter \(T_1\) values in samples with high spin concentration, whereas the opposite trend is observed at low temperatures. Furthermore, these results demonstrate the ability to tune the \(T_1\)/\(T_m\) ratio from a few hundred to a few hundred thousand (see ESI). This is particularly important in quantum memory applications, where microwave pulses and quantum states are stored as partially excited spin states which thermalise on a \(T_1\) timescale. The high \(T_1\)/\(T_m\) ratio indicates that a high number of states can be stored for long periods without loss or processing errors induced by thermalisation.

To explore the potential limits of our system for storing quantum coherent states, we employed a dynamical decoupling protocol at RT and on a sample of 0.1\% BDPA in OTP according to the Carr-Purcell-Meiboom-Gill (CPMG) sequence (Fig.~\ref{fig:fig2}(d)). These sequences are capable of extending the effective \(T_m\) by partially decoupling electron spins from the surrounding nuclear spin bath. Bonizzoni \textit{et al.,} recently demonstrated the impact of CPMG on BDPA at low temperatures to prolong \(T_m\) and improve the sensitivity of BDPA as a magnetic field sensor\cite{Bonizzoni2024}. Here, the sequence was optimised by varying the length of the interpulse delay, \(\tau\), enabling the application of up to 70 pulses to extend \(T_m\) (termed \(T_{DD}\)). \(T_{DD}\) was determined by fitting the extended decay trace to a monoexponential function, revealing an extended effective coherence up to 11.5 \(\mu\)s, similar to the values achieved by Bonizzoni \textit{et al.,} at 3 K using a coplanar waveguide and 20 refocusing \(\pi\) pulses\cite{Bonizzoni2024}.

\subsection{Probing two-level quantum dynamics at room temperature}

\begin{figure}[!b]
  \centering
  \includegraphics[width=0.8\linewidth]{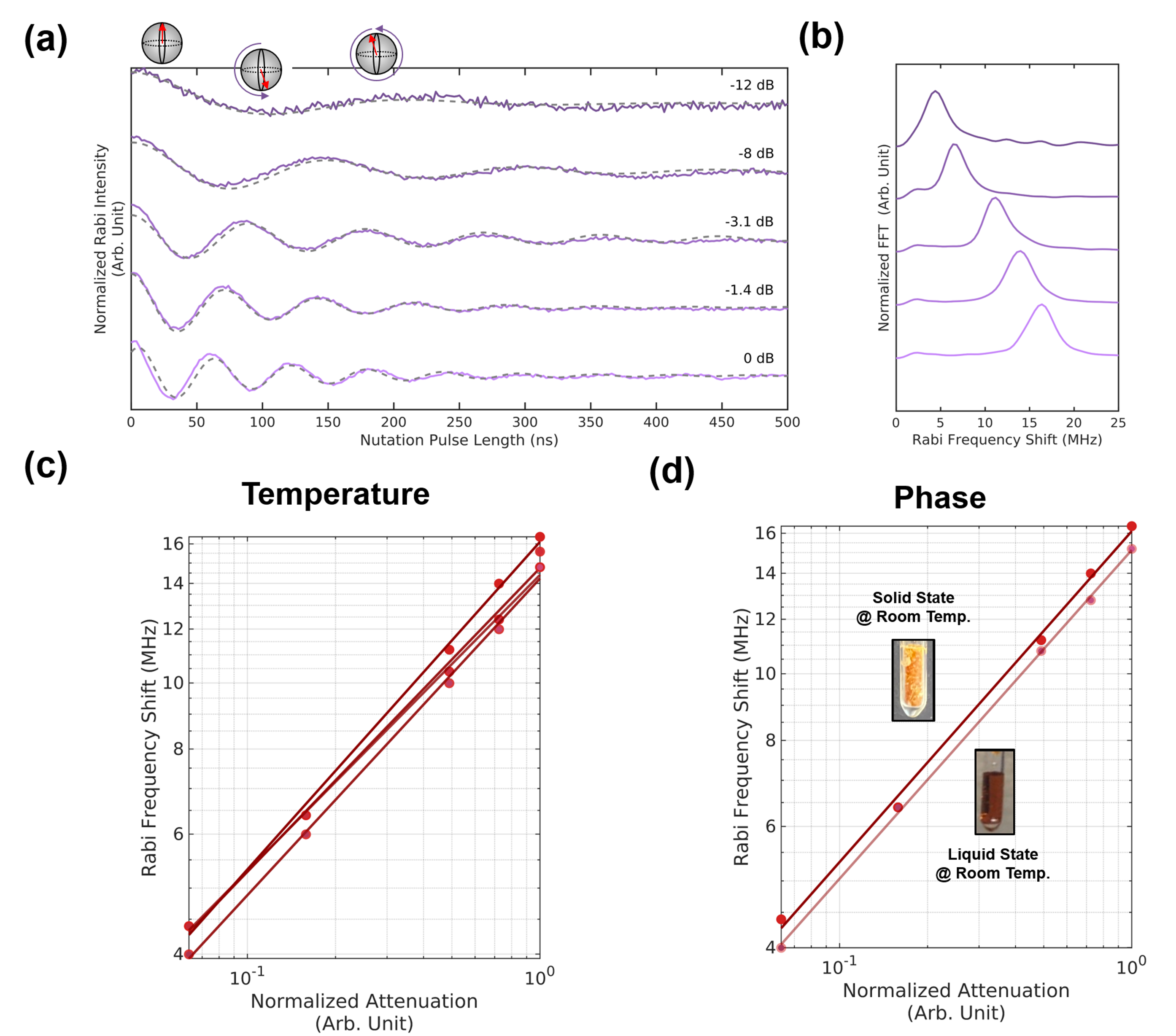}
  \caption{Rabi Oscillation Experiment for 0.1\% BDPA over variable temperatures. (a) Nutation curves with normalized Rabi intensities as a function of nutation pulse length at various microwave attenuation levels measured at RT. The data points represent experimental measurements, while the grey dashed lines indicate the fits to a damped sinusoidal function, providing a precise determination of the Rabi frequencies. Insets illustrate the corresponding spin rotations on the Bloch sphere. (b) Shows the Rabi frequency shifts across a spectrum of microwave powers, reflecting the system's resonance characteristics. (c) and (d) plot the Rabi frequency shift against normalized microwave attenuation (arbitrary units), categorized by temperature (5K, 70K, 150K, 200K, and 300K from light to dark) and phase (liquid and solid state), respectively. The fitting curves illustrate the model's alignment with the experimental data, accentuating the temperature's role in modulating electron spin coherence.}
  \label{fig:fig3}
\end{figure}

Next, to determine whether this spin system would be suitable as a two-level quantum system, we conducted Rabi oscillation experiments under different temperatures, phases, and microwave powers, as shown in Fig~\ref{fig:fig3}(a). Ideally, for quantum memory based on a cavity quantum electrodynamics (cQED) approach, a spin ensemble should exhibit strong spin-photon coupling to store quantum states for prolonged periods efficiently. The threshold for strong spin-photon coupling can be effectively described by a system's "cooperativity", \(C = 4g_e^2 / (\kappa_c \kappa_s)\), where \(g_e\) is the spin-photon coupling strength, \(\kappa_c\) is the cavity-photon decay rate, and \(\kappa_s = 2 / T_2^*\)\cite{Breeze2017}. Here, \(\kappa_c = 2 \pi f_{mode} / Q_L\), where \(Q_L\) is the loaded quality factor of the resonator ($\approx 200$), and \(f_{mode}\) is the microwave frequency ($\approx 9.65$ GHz), giving a \(\kappa_c\)-value $\approx300$ MHz. In strong spin-photon coupling systems where \(C\)>1, \(g_e\) is often taken as \(\Omega / 2\), where \(\Omega\) is the Rabi oscillation frequency.

According to our measurements using a sample at 0.1\% BDPA concentration at RT, the Rabi frequency ranges between 4.1 and 16.5 MHz depending on the applied microwave power (Fig.~\ref{fig:fig3}(a) and (b)). While this demonstrates the ability to coherently drive BDPA as a two-level quantum system, the Rabi frequency is below the strong coupling threshold, which is confirmed by the proportional dependence of the Rabi oscillation frequency on the microwave power (Fig.~\ref{fig:fig3}(b)). While we were unable to directly measure \(T_2^*\) in our experiments, an estimate can be made from the full-width half maximum (FWHM) of the inhomogeneous linewidth (=1/$\pi$\(T_2^*\)) measured from the steady-state radical resonance (FWHM $\approx$ 22 MHz $\approx$ 14 ns, Fig.~\ref{fig:fig2}(a)). From this, and using \(g_e\) = 8 MHz, then C $\approx 0.01$. Improving to \(C > 1\) would thus require improvements in \(Q_L\), reduction in \(f_{mode}\), or, most significantly due to the \(g_e^2\) term, an improvement in the spin-photon coupling strength. For example, critically coupled sapphire resonators with \(Q_L\) >10,000 are common in X-band EPR spectroscopy and could offer a potential factor of 50 improvement in \(\kappa_c\). Furthermore, assuming that all spins exhibit an equal individual coupling strength, then \(g_e\) $\approx$ \(g_s \sqrt{N}\), where \(g_s = \gamma \sqrt{\mu_0 \hbar f_{mode} \pi / V_m}\). Here, \(\gamma\) is the electron gyromagnetic ratio ($\approx 28.024$ GHz/T for an organic radical), \(\mu_0\) is the permeability of free space, \(\hbar\) is the reduced Planck constant, and \(V_m\) is the resonator's magnetic mode volume. From this relationship, it becomes apparent that the principal routes to improving \(g_e\) are increasing \(\sqrt{N}\), reducing \(f_{mode}\) or \(V_m\). To reduce \(V_m\), it would be necessary to design a new microwave resonator using materials with a significantly higher dielectric permittivity (\(\epsilon_r\)), since \(V_m\) scales by \(\epsilon_r^{-3/2}\). This approach has been successful in the field of maser research, where switching from a sapphire (\(\epsilon_r = 9\)) to strontium titanate (STO, \(\epsilon_r = 318\)) resonator was used to generate Dicke states at RT using photoexcited pentacene or 6,13-diazapentacene-doped p-terphenyl\cite{Breeze2017, Ng2023}. Furthermore, Lenz \textit{et al.,} have measured a state of strong coupling for neat samples of 2,2-diphenyl-1-picrylhydrazyl (DPPH) and BDPA at RT using Fabry-Perot resonators at Q-band, despite each radical exhibiting significantly lower \(T_m\) values compared to our system\cite{Lenz2019, Lenz2020, Lenz2021}. We note that the high \(\epsilon_r\) approach would also enable six-fold or more miniaturization in the resonator's size due to near-total confinement of the electric and magnetic fields of the \(TE_{01\delta}\) mode compared to a sapphire resonator.

Our Rabi oscillation experiments were performed at various microwave attenuations (Fig.~\ref{fig:fig3}(a) and (b)) and temperatures (see ESI). To fit the Rabi oscillations, we use a damped sinusoidal function given by: \(I(t) = A e^{-\lambda t} \sin(2 \pi \Omega_R t + \phi) + s\), where \( I(t) \) is the measured intensity as a function of time \( t \), \( A \) is the amplitude of the oscillation, \( \lambda \) is the damping rate, \( \Omega_R \) is the Rabi frequency, \( \phi \) is the phase of the oscillation, and \( s \) is the offset. For microwave attenuations of 0 dB (baseline), 1.4 dB, 3.1 dB, 8 dB, and 12 dB, we obtained Rabi frequencies  \( \Omega_R \) of 16.5, 14.2, 10.3, 6.1, and 4.1 MHz at RT, respectively. The linear relationships shown in Fig.~\ref{fig:fig3} (c) and (d) confirm that the oscillations are of pure Rabi type\cite{moreno2020molecules} and demonstrate the potential to drive BDPA as quantum spin system. According to Maylander \textit{et al.}\cite{mayländer2023room}, the number of single-qubit logic operations \(\Omega_M\) can be calculated using \(\Omega_M = 2 T_m \Omega_R\). In our experiments, 0.1\% BDPA in OTP at RT achieves a value over 400, which is 20 times higher than the previous molecular-spin-qubit research. Furthermore, these experiment highlight the remarkable thermal stability of solid-state BDPA-doped OTP, demonstrating consistent performance across a temperature range of 5K to 300K. This stability is evidenced by the Rabi oscillation frequencies and nutation curves, which remain largely invariant throughout this extensive temperature spectrum. Furthermore, Fig.~\ref{fig:fig3}(d) highlights the significant phase stability of BDPA-doped OTP at RT, as consistent Rabi oscillation frequencies are observed in both solid and liquid phases. This uniform behaviour across different phases underscores the material's robust coherence properties, which are crucial for developing reliable quantum information processing (QIP) systems. Notably, the transition between solid and liquid states does not result in appreciable dephasing, thereby preserving the integrity of qubit systems.

\subsection{Generation of radical spin polarisation using photoexcited triplets}

To further explore the versatility of BDPA as a platform for quantum applications, we sought to introduce mechanism to introduce a photoexcited spin polarisation state using the electron spin polarisation transfer mechanism (ESPT) and radical-triplet pair mechanism (RTPM) which was previously explored by Blank and Levanon for the BDPA radical\cite{Blank2001, Blank20012b}. These reports established H$_{2}$TPP and BDPA as a promising system for quantum amplifiers due to the system's remarkable ability to induce a spin-polarised state lasting >100 $\mu$s in vicious solutions of 1-chloronaphthalene and paraffin. However, we noted that the \(T_1\) ($\approx$ 1.2 ms measured for BDPA in OTP was significantly longer than the \(T_1\) = 6.9 $\mu$s measured in their work, and thus postulated that this would result in a significantly longer spin polarisation. We prepared samples by dissolving powders of H$_{2}$TPP and BDPA in OTP at 0.1$\%$ total spin concentration by melting the mixture at 60$^{o}$C and allowing it to cool. Samples were then measured by time-resolved (tr)EPR spectroscopy using pulsed photoexcitation at 650 nm to generate H$_{2}$TPP triplet states. 

\begin{figure}[!b]
  \centering
  \includegraphics[width=0.6\linewidth]{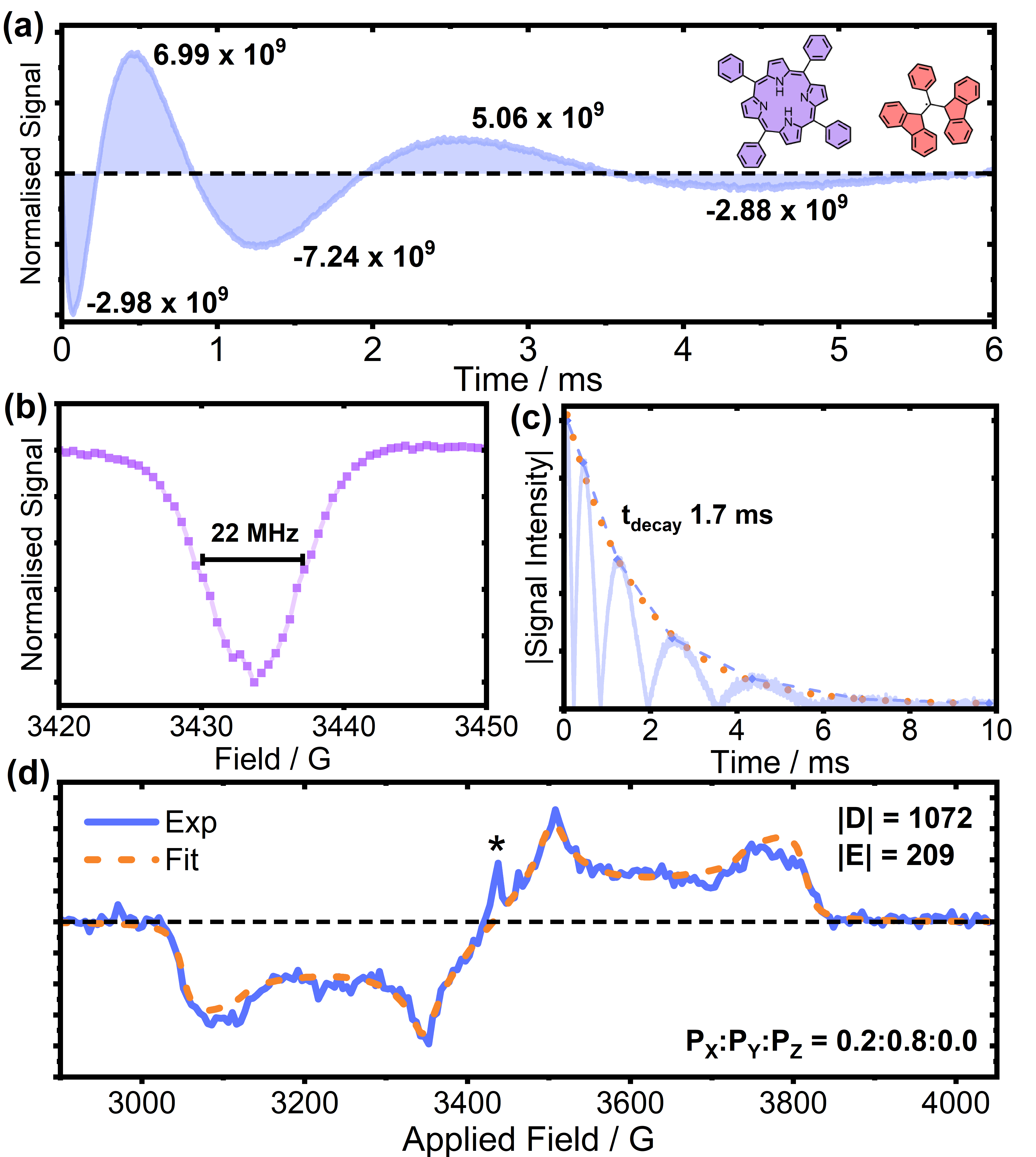}
  \caption{Photoexcited time-resolved EPR spectra for a mixture of 0.1$\%$ H$_{2}$TPP and BDPA in OTP, showing (a) triplet-radical pair spin polarisation oscillations over 6 ms; (b) field swept transient of the centrefield radical 160 \(\mu\)s after the laser flash; (c) spin polarisation decay of the absolute signal intensity and; (d) powder phase photoexcited EPR spectrum of H$_{2}$TPP 440 ns after laser flash. Zero-field splitting values given in MHz.}
  \label{fig:fig4}
\end{figure}

Remarkably, these measurements revealed a pronounced oscillating spin polarisation with an overall net emissive gain (Fig.~\ref{fig:fig4}(a) and (b)). The sequential emissive and absorptive phases peak at 0.15, 0.5, 1.25, 2.5 and 4.5 ms, lasting 0.25, 0.6, 1.1, 1.6 and 3.3 ms, respectively. The absolute polarisation reduces monoexponentially with a decay lifetime (t$_{decay}$) of 1.7 ms (Fig.~\ref{fig:fig4}(c)), which is slightly longer than the \(T_1\) measured for BDPA. Notably, these oscillations were not present in the solid state following vitrification of the OTP and are also dependent on the sample's composition (see ESI). These findings are consistent with both ESPT and RTPM which each require solution phase physical interactions between the radical and triplet moieties to induce spin polarisation. Interestingly, we further find that in the solid state, the sign of the radical polarisation is absorptive rather than emissive. Since the sign of the polarisation of determined by the multiplicity and exchange coupling of the interacting species, this could suggest a different polarisation mechanism. The triplet signal measured in the solid state could be fitted to S = 1 Hamiltonian and found to agree with zero-field splitting values and triplet sub-level populations typically observed for free-base porphyrin\cite{Kay2003, Yabuki2023} (Fig.~\ref{fig:fig4}(d)). Attempts to measure the intermediate quartet state which considered to be responsible for generating an emissive radical in ESPT and RTPM failed. A much more comprehensive analysis of this new phenomenon will form the basis of future work.

\section{Discussion}

Our results demonstrate that organic radicals in a suitable host matrix, in this case, BDPA in OTP, can harbour robust spin properties applicable to a range of quantum technologies including quantum memory, and potentially even quantum amplifiers and spin refrigerators. By employing OTP as a viscous liquid host, we show it is possible to exhibit remarkably long \(T_1\) while also protecting electron spins from dipole coupling and thermally induced decoherence effects, both of which can be improved by employing more dilute spin mixtures or using dynamical decoupling protocols. We further demonstrate the capacity of BDPA act as a coherently manipulable system at RT close to the strong coupling limit, as well as its compatibility with spin polarisation methods by the inclusion of a photo-excitable triplet moiety to realise a rich variety of unique quantum behaviours.  

\begin{figure}[!b]
  \centering
  \includegraphics[width=0.6\linewidth]{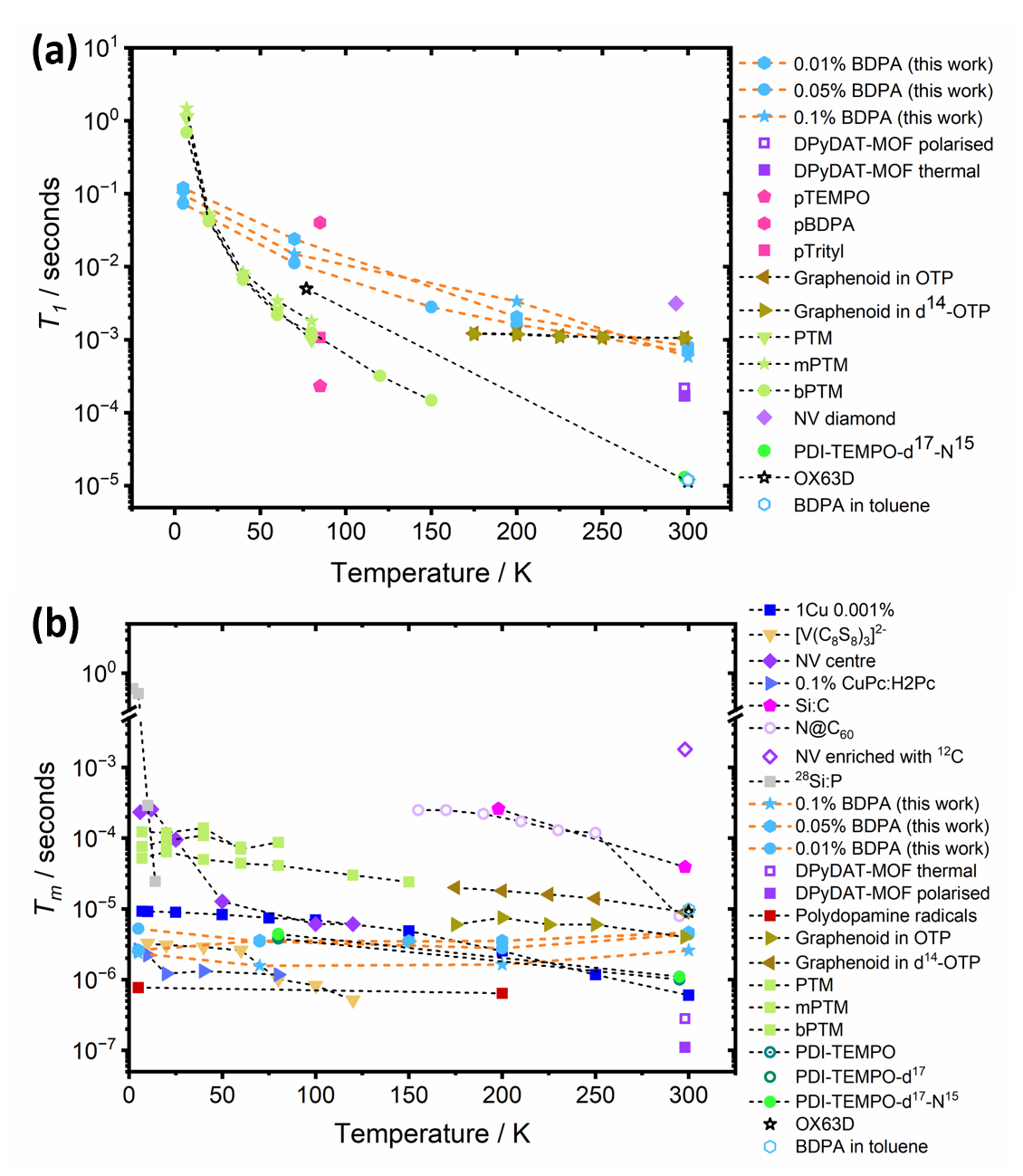}
  \caption{Comparison of (a) \(T_1\) and (b) \(T_m\) for  cutting-edge spin-qubit systems for QIP. The data for each spin system are sourced from 1Cu 0.001\%\cite{Bader2014}, [V(C$_8$S$_8$)$_3$]$^{2-}$\cite{zadrozny2015millisecond}, NV center\cite{takahashi2008quenching}, 0.1\% CuPc:H$_2$Pc, SiC\cite{koehl2011room}, N@C$_{60}$\cite{morton2006electron}, Fe$_8$\cite{takahashi2011decoherence}, NV center in $^{12}$C-enriched diamond\cite{balasubramanian2009ultralong}, Pn-MOF\cite{yamauchi2024room}, $^{28}$Si:P\cite{tyryshkin2012electron}, mPTM radical\cite{schafter2023molecular}, DPyDAT-MOF \cite{Orihashi2023}, polydopamine radicals \cite{Tadyszak2021}, pTEMPO, pBDPA, pTrityl \cite{Avalos2020}, graphenoid in OTP \cite{Lombardi2022}, PTM, mPTM, bPTM \cite{Schafter2023}, PDI-TEMPO \cite{mayländer2023room}, BDPA in toluene \cite{Meyer2014}, OX63D \cite{chen2016electron}. It is important to note that the data included in this comparison were not all collected under identical magnetic field conditions, and for instances where a bi-exponential decay was reported, the slow decay constant has been used when available.}
  \label{fig:fig5}
\end{figure}

Molecular electron spins are subject to various burdensome decoherence sources, including nuclear spins, phonons, and intermolecular electron dipole interactions, which have effectively been probed during this investigation. The BDPA radical is stabilised by delocalisation of spin density over the fluorene groups, which also effectively acts to partially homogenise the local spin bath.\cite{Haeri2016, Levien2021} In the dilute and fast-motion regime, BDPA measured in \(h^8\)-toluene at room temperature is known to exhibit an impressive \(T_m\) of 9.8 $\mu$s, limited by local mode relaxation.\cite{Meyer2014} However, in the solid-state or slow-motion regimes (as is the case here), one observes a resolved EPR resonance structure resulting from hyperfine coupling to four inequivalent protons that comprise the dominant source of electron spin decoherence. It is also notable that our spin systems appear to be close to (for 0.1\%), or below (for 0.05 and 0.01\%), the concentration limit for significant intermolecular dipole or exchange interactions. This is initially indicated by the similarity of \(T_1\) measured for each spin concentration and also supported by comparison with data reported by Li \textit{et al.,} wherein BDPA was measured in partially deuterated OTP at X- and Q-band frequencies.\cite{Li2020} In their study, and at concentrations two to eight times higher than our samples, the lineshape of BDPA exhibits broadened field-swept echos attributed to electron-electron interactions that become absent at low concentrations, and yield lineshapes similar to our measurements. Notably, both the dilution of BDPA from 0.1\% to 0.05\%, and the employment of dynamical decoupling pulses significantly extended \(T_m\), while the impact of the former on \(T_1\) was less significant. Furthermore, the clear dependence of \(T_1\) on temperature is consistent with the presence of phonon-mediated relaxation below RT, however, due to the length of \(T_1\) compared to \(T_m\), these effects did not appear to significantly impact \(T_m\). Altogether, these findings are consistent with the main contributors to decoherence in BDPA being derived from electron spin dipole interactions and the nuclear spin bath. Indeed, relatively recent findings regarding organic radicals have predicted that proton spins up to 12 angstroms from the electron spin centre can contribute to decoherence\cite{Canarie2020}. As a result, extending \(T_m\) has often required impractical but nuclear spin-free solvents, like CS\(_{2}\), or synthesis of novel molecules with reduced nuclear spin noise\cite{Graham2017, Gaita-Arino2019}. Here, OTP is demonstrated to be a versatile and convenient alternative by enabling remarkable spin relaxation and decoherence parameters for BDPA (Fig.~\ref{fig:fig5}(a) and (b)), while also enabling other useful functionalities, such as adjustable composition and the ability to easily include photoactivated auxiliary molecules. These insights also underscore that, unlike with other solid-state inorganic spin defect systems based in diamond,\cite{Liu2018, Park2022} silicon carbide,\cite{Nagy2019} or most recently, hexagonal boronitride,\cite{Stern2022, Guo2023, Stern2024} molecular spin systems offer on-demand control over both temperature and dopant concentration can optimize \(T_1\) and \(T_m\), offering pivotal guidelines for designing materials with tailored coherence times suited for quantum memory applications.

The ability of BDPA-doped OTP to preserve quantum coherence from cryogenic to ambient conditions positions it as an adaptable material for QIP technologies. Our Rabi oscillation experiments revealed the remarkable thermal and phase stability of BDPA-doped OTP and strongly suggest the ability to conduct over 400 coherent manipulations even at RT, significantly surpassing previous molecular benchmarks\cite{mayländer2023room}  
To realise strong spin-photon coupling for efficient quantum memory and sensing applications it would be necessary to increase the Q$_f$ of the resonator, increase the resonant frequency, increase spin concentration, and/or improve \(T_2^*\). Lenz \textit{et al.,} successfully demonstrated strong spin-photon coupling at Q-band by employing a Fabry-Perot resonator and pellets of the BDPA-benzene complex, and separately, DPPH\cite{Lenz2019, Lenz2021}. Importantly, these systems exhibit significantly lower \(T_m\), and by extension, likely \(T_2^*\), which presents an opportunity for spin systems with longer \(T_m\)s to demonstrate strong spin-photon coupling while maintaining relatively low Q$_f$ or spin concentrations.

Applications of BDPA towards a range of quantum technologies often require an initialisation step. For example, for QIP it is necessary to obtain a "pure" quantum state with a known two-level spin population, and similarly for quantum amplifiers based on masers, an emissive spin polarisation is required. Our demonstration of inducing spin polarisation using photoexcitation combined with triplet-radical interactions affirms the versatility of molecular media for quantum applications. Furthermore, we note the strong and long-lived oscillations in spin polarisation could be useful for the development of (liquid phase) ultra-low noise amplifiers,\cite{Arroo2021} but could also be an enabling technology for high-temperature quantum technologies by acting as a microwave mode cooler\cite{Wu2021, Ng2021, Blank2023}. Previous work by Blank and Levanon demonstrated that high-viscosity solvents facilitated longer spin polarization lifetimes \cite{Blank2001, Blank20012b}. Hence, for our investigations, the RT glass phase of OTP was ideal due to its relatively high viscosity (\(\eta \approx 8\) mPa s at 300 K)\cite{Casalini2016} and optical clarity. Organic glass hosts without appreciable flow properties, such as 1,3,5-tri(1-naphthyl)benzene\cite{Schroder2022, Attwood2023} or \(\beta\)-estradiol~\cite{35,36}, were not suitable since they prohibit transient short-range interactions required for ESPT and RTPM. While the origin of the oscillations has yet to be completely understood by our group, it could reasonably stem from extended contributions from RTPM resulting in further polarisation exchanges between the H$_{2}$TPP and BDPA. These intermolecular interactions on a millisecond time scale would be enabled by the long (\(\approx 6\) ms) triplet lifetime of H$_{2}$TPP \cite{Gouterman1974}, the long \(T_1\) of BDPA, and the inefficiency of triplet quenching associated with RTPM\cite{BLANK2002}. A key objective moving forward is to explore this regime further to understand the mechanism, and hopefully tune the behaviour towards particular applications.

\section{Conclusion and outlook}

An important objective in the development of quantum technologies is to develop alternative tunable qubit media to enable robust quantum activity and manipulation over a range of temperatures. Here, by employing known chemical strategies to facilitate robust quantum behaviours, we have demonstrated remarkably long \(T_1\) and \( T_m\) times for the well-known BDPA organic radical system, which compares favourably against state-of-the-art molecular qubit media (Fig.~\ref{fig:fig5}(a) and (b)), and established behavioural regimes applicable to several emerging quantum technologies. Moreover, we anticipate that by capitalising on the structural versatility enabled by chemical synthesis these remarkable quantum functionalities can be significantly improved upon. There are now several radical-chromophore systems\cite{39,40,41, Horwitz2016}, some with already long-lived \(T_1\) and \( T_m\) that could be improved at high temperatures simply by employing the appropriate host matrix. More vigorous strategies could also include deuteration \cite{Delage-Laurin2021} to improve \( T_m\) and \(T_2^*\), or the use of alternative radicals \cite{Schafter2023, Lombardi2022, Quintes2023} and alternative pulsed protocols\cite{Bonizzoni2024}. We note, however, that in addition to the realisation of strong spin-photon coupling, additional design criteria are required in the realistic scenario for quantum computation. Most notably, spin qubits must be deployed in a controllable host matrix that enables intermolecular entanglement through dipole coupling or exchange interactions, whilst being compatible with appropriate substrates.\cite{Zhou2024} This is perhaps the most significant obstacle facing QIP-relevant molecular qubit technology due to the challenging synthetic chemistry involved, the employment of an appropriate deposition approach, and compensating for the reduced coherence associated with coupled paramagnetic states. However, the incorporation of a robust entanglement faculty in spin-ensemble quantum memories can also enable their application in quantum network and communications technology.

\section*{Acknowledgments}
This work was supported by the U.K. Engineering and Physical Sciences Research Council through Grants No. EP/W027542/1. We would also like to acknowledge funding for our SPIN-Lab EPR facility funded by the EPSRC (EP/P030548/1). The authors would also like to thank Nobuhiro Yanai (Kyushu University, Japan) and Vinayak Rane (Indian Institute of Geomagnetism, India) for helpful discussions regarding our time-resolved EPR results.

\section*{Conflict of Interest}
The authors declare no conflict of interest.

\bibliographystyle{abbrv}  
\bibliography{references} 

\end{document}